\documentclass[letterpaper, conference, 10 pt]{ieeeconf}
\IEEEoverridecommandlockouts
\usepackage[english]{babel}

\usepackage{amsmath, amssymb, amsthm}
\usepackage{cite}
\usepackage{xcolor}
\usepackage{caption}
\usepackage{subcaption}
\usepackage[linesnumbered,ruled,vlined]{algorithm2e}
\usepackage{optidef}
\let\labelindent\relax

\usepackage{enumitem}
\usepackage{url}
\usepackage{accents}

\def\matt#1{\begin{bmatrix}#1\end{bmatrix}}

\def\BibTeX{{\rm B\kern-.05em{\sc i\kern-.025em b}\kern-.08em
		T\kern-.1667em\lower.7ex\hbox{E}\kern-.125emX}}

\title{\LARGE \bf
	Variable Sampling MPC via Differentiable Time-Warping Function
}

\author{Zehui Lu, \ Shaoshuai Mou
	\thanks{The authors are with the School of Aeronautics and Astronautics, Purdue University, IN 47907, USA {\tt\small \{lu846,mous\}@purdue.edu} }
	\thanks{This work was supported by a gift funding from Northrop Grumman Corporation.}
}

\begin{document}

\maketitle
\thispagestyle{empty}
\pagestyle{empty}

\begin{abstract}
	Designing control inputs for a system that involves dynamical responses in multiple timescales is nontrivial. This paper proposes a parameterized time-warping function to enable a non-uniformly sampling along a prediction horizon given some parameters. The horizon should capture the responses under faster dynamics in the near future and preview the impact from slower dynamics in the distant future. Then a variable sampling MPC (VS-MPC) strategy is proposed to jointly determine optimal control and sampling parameters at each timestamp. VS-MPC adapts how it samples along the horizon and determines optimal control accordingly at each timestamp without offline tuning or trial and error. A numerical example of a wind farm battery energy storage system is also provided to demonstrate that VS-MPC outperforms the uniform sampling MPC.
\end{abstract}


\section{Introduction}\label{sec:introduction}
In many applications such as energy management systems, transportation, aerospace systems, and process control systems, a primary task is to make real-time decisions or scheduling while optimizing a specific objective and not violating some constraints \cite{lee2011model}.
If a dynamical model of such a system is available prior, one commonly used method is model predictive control (MPC)
\cite{lee2011model}.
To formulate an MPC problem, a discrete-time dynamical model of the system is typically required to construct a discrete-time prediction of system behaviors over a specific prediction horizon.
To achieve an optimum of a given objective, MPC methods usually explore all possible control inputs while guaranteeing that these control inputs can forward propagate the given dynamics correctly and not violate any given constraints.

When MPC is applied to multi-timescale systems,
such as power grids\cite{clarke2018hierarchical}, chemical processes \cite{wei2017multi}, aerospace systems\cite{koeln2019hierarchical,wang2020hierarchical}, and electrified vehicles\cite{Huazhen2022ACC, amini2020hierarchical, hu2021multihorizon},
the slower dynamics often require a longer prediction horizon. This usually leads to a higher-dimension MPC problem with a larger computational burden. To address this challenge, the singular perturbation theory\cite{kokotovic1999singular} has been explored broadly, which decomposes a multi-timescale system into two subsystems with faster and slower timescales, respectively.
Then an MPC controller can be developed for each of these two subsystems.
But this method is only applicable to the systems whose dynamics can be explicitly decomposed into two subsystems with faster and slower timescales.

Another method of controlling multi-timescale systems is hierarchical MPC (H-MPC) \cite{koeln2019hierarchical,clarke2018hierarchical,wang2020hierarchical,amini2020hierarchical,farina2018hierarchical,pangborn2020hierarchical}. The H-MPC method first computes an optimal reference by an MPC given slower dynamics over a relatively long prediction horizon. Then, a reference tracking problem is solved by another controller given faster dynamics over a shorted prediction horizon and hence some optimal control inputs can be obtained.
Besides time delays arising from communication among controllers, choosing a proper quantity as the reference value requires prior knowledge of the specific system.

Briefly, to make MPC able to deal with multi-timescale systems well, the prediction horizon should be longer to capture more look-ahead information in the distant future yet the sampling rate of MPC should be small enough to provide more accurate prediction in the near future.
A multi-horizon MPC (MH-MPC) \cite{hu2021multihorizon,hu2022spatial} has been studied recently, which combines a short receding horizon and a long shrinking horizon altogether in one MPC formulation. The short receding horizon indicates relatively accurate prediction with a higher sampling rate, whereas the long shrinking horizon extends to the end of the trip with a lower sampling rate. Even though MH-MPC exploits preview information over a longer horizon, it introduces an extra computational burden, especially at the beginning of the trip, because the dimension of the MH-MPC problem varies and depends on the current progress over the entire trip. In addition, MH-MPC requires the entire trip to be finite.
Another direction to develop MPC for multi-timescale systems is by the non-uniform sampling MPC (NS-MPC) \cite{tan2016model,gomozov2016adaptive,liao2017cascaded,brudigam2021model}, in which the prediction horizon is partitioned into multiple parts and each part has a different sampling rate. The dimension of the decision variables, i.e., the number of prediction steps, is assumed to be fixed to avoid an extra computational footprint.
Both NS-MPC and MH-MPC require manual tuning of some parameters to obtain better performance by trial and error. And determining the optimal settings involves trial and error, and requires expert knowledge of the specific system.

Instead of manually tuning some parameters of a prediction horizon, this paper seeks a differentiable temporal mapping from sampling time to actual time such that it can describe any non-uniform sampling under some parameterizations.
Time-warping functions represent this kind of mapping, which was originally proposed to deal with the time misalignment between two temporal sequences\cite{sakoe1978dynamic}, or between human demonstrations and system observations \cite{kingston2011time, jin2022learning}.
To describe the faster dynamics in the near future precisely and preview the impact of the slower dynamics in the distant future, there are some constraints on a time-warping function, which require the function's differentiability. Then the function's parameters can be a part of decision variables to be optimized while designing control inputs at run-time.

To control a multi-timescale system with one MPC controller and avoid the manual tuning of sampling, this paper proposes a variable sampling MPC (VS-MPC) strategy to accurately capture the responses under the faster dynamics in the near future and preview the impact from the slower dynamics in the distant future.
In detail, a differentiable time-warping function describes the timeline of a prediction horizon.
The function is parameterized by some decision variables and an optimal control problem jointly determines the control inputs and function parameters at each timestamp without any manual tuning on the horizon. As the situation changes at run-time, VS-MPC adapts how it samples along the horizon and then determines optimal control accordingly.
In addition, Section \ref{sec:simulation} studies how the proposed VS-MPC strategy performs in a specific application where a control strategy needs to be designed to control a battery energy storage system (BESS) for a wind farm. A performance comparison for several methods is also included in Section \ref{sec:simulation}.


\emph{Notation } The real number set is $\mathbb{R}$. The non-negative real number set is $\mathbb{R}_{\geq 0}$.
Let $\text{col}\{ \boldsymbol{v}_1, \cdots, \boldsymbol{v}_a \}$ denote a column stack of elements $\boldsymbol{v}_1, \cdots, \boldsymbol{v}_a $, which may be scalars, vectors or matrices, i.e. $\text{col}\{ \boldsymbol{v}_1, \cdots, \boldsymbol{v}_a \} \triangleq {\matt{{\boldsymbol{v}_1}^{\prime} & \cdots & {\boldsymbol{v}_a}^{\prime}}}^{\prime}$.
For a scalar $x \in \mathbb{R}$, $[x]^{+} \triangleq x$ when $x \geq 0$ and 0 otherwise.
$\mathcal{N}(\mu, \sigma^2)$ indicates a normal distribution with a mean $\mu$ and a standard deviation $\sigma$.

\section{Problem Formulation}\label{sec:problem}

Suppose the continuous-time open-loop system dynamics for a plant are described by
\begin{equation} \label{eq:continuous_dyn}
	\dot{\boldsymbol{x}}(t)=\boldsymbol{f}_c(\boldsymbol{x}(t), \boldsymbol{u}(t)),
\end{equation}
where $t \in \mathbb{R}_{\geq 0}$ denotes time, $\boldsymbol{x}(t) \in \mathbb{R}^{n}$ denotes state at time $t$,  $\boldsymbol{u}(t) \in \mathbb{R}^{m}$ denotes input at time $t$, and $\boldsymbol{f}_c:\mathbb{R}^{n}\times\mathbb{R}^{m}\times\mathbb{R}^r\mapsto\mathbb{R}^{n}$ denotes the nonlinear dynamics.

The open-loop control $\boldsymbol{u}(t)$ is determined by discrete-time model predictive control with sampling in the following way.
Let $N$ denote the number of steps in a prediction horizon and $\boldsymbol{x}_k$ denote the value of $\boldsymbol{x}(t)$ at the sampling time $t_k$, $k=0, 1, 2, \cdots$, i.e. $\boldsymbol{x}_k = \boldsymbol{x}(t_k)$.
By Euler integration with non-uniform sampling time $\Delta_k > 0$ at time $t_k$, one reaches the following discretization of the continuous system in \eqref{eq:continuous_dyn}:
\begin{equation}
	\boldsymbol{x}_{k+1} = \boldsymbol{x}_{k} + \Delta_k \boldsymbol{f}_c(\boldsymbol{x}_k, \boldsymbol{u}_k).
\end{equation}
Let 
$$J \triangleq J(\boldsymbol{x}_{0:N|k},\boldsymbol{u}_{0:N-1|k}, \Delta_{0:N-1|k}),$$
where $\boldsymbol{x}_{0:N|k} \triangleq \text{col}\{\boldsymbol{x}_{k},\boldsymbol{x}_{k+1|k}\cdots,\boldsymbol{x}_{k+N|k}\} \in \mathbb{R}^{n(N+1)}$ denotes the state at current time $t_k$ and the states from the future time $t_{k+1}$ to $t_{k+N}$ that are predicted at time $t_k$; similarly $\boldsymbol{u}_{0:N-1|k} \triangleq \text{col}\{\boldsymbol{u}_{k|k},\cdots,\boldsymbol{u}_{k+N-1|k}\} \in \mathbb{R}^{mN}$; $\Delta_{0:N-1|k} \triangleq \text{col}\{\Delta_{k|k},\cdots,\Delta_{k+N-1|k}\} \in \mathbb{R}^{N}$ denotes the non-uniform sampling time intervals from time $t_k$ to $t_{k+N-1}$ that are determined at current time $t_k$. Note that for a uniform sampling, $\Delta_k$ is a constant for any time $t_k$.

Then at time $t_k$, given the current state $\boldsymbol{x}_{k}$, the optimal control can be determined by
\begin{mini}|s|
	{\substack{\boldsymbol{u}_{0:N-1|k} \\ \Delta_{0:N-1|k}}}{ J(\boldsymbol{x}_{0:N|k},\boldsymbol{u}_{0:N-1|k}, \Delta_{0:N-1|k}) \label{oc}}
	{}{}
	\addConstraint{ \boldsymbol{x}_{k+j+1|k} = \boldsymbol{x}_{k+j|k} + \Delta_{k+j|k} \boldsymbol{f}_c(\boldsymbol{x}_{k+j|k}, \boldsymbol{u}_{k+j|k})}
	\addConstraint{\forall j=0,\cdots,N-1 \  \text{with given } \boldsymbol{x}_{k}}
	\addConstraint{\boldsymbol{g}(\boldsymbol{x}_{0:N|k}, \boldsymbol{u}_{0:N-1|k}, \Delta_{0:N-1|k}) \leq \boldsymbol{0}}
	\addConstraint{\boldsymbol{h}(\boldsymbol{x}_{0:N|k}, \boldsymbol{u}_{0:N-1|k}, \Delta_{0:N-1|k}) = \boldsymbol{0},}
\end{mini}
where $\boldsymbol{g}(\boldsymbol{x}_{0:N}, \boldsymbol{u}_{0:N-1}, \Delta_{0:N-1|k})$ denotes a column stack of inequality constraints; $\boldsymbol{h}(\boldsymbol{x}_{0:N}, \boldsymbol{u}_{0:N-1}, \Delta_{0:N-1|k})$ denotes a column stack of equality constraints; $\leq$ and $=$ in these constraints indicate element-wise inequality and equality.
The discrete-time optimal control determined at time $k$ will be denoted by $\boldsymbol{u}^*_{0:N-1|k}$.
Then in a receding horizon fashion, the system will perform the optimal control $\boldsymbol{u}^*_{k|k}$ at time $t_k$, update its states at time $t_{k+1}$, and then rerun the optimal control problem \eqref{oc} with current state $\boldsymbol{x}_{k+1}$. This procedure will be performed repeatedly under a prescribed frequency.
Solving \eqref{oc} repeatedly in a receding-horizon fashion at run-time is unrealistic because determining the sampling rates $\Delta_{0:N-1|k}$ requires offline manual tuning by trial and error for every specific application.
The \textbf{problem of interest} is to find the discrete-time optimal control $\boldsymbol{u}^*_{0:N-1|k}$ and sampling steps $\Delta_{0:N-1|k}$ jointly at each time $t_k$, without any manual tuning of $\Delta_{0:N-1|k}$ afterward.

\section{Approach}\label{sec:approach}
This section introduces a variable sampling MPC (VS-MPC) strategy, which partitions a prediction horizon with non-uniform sampling by a time-warping function. The time-warping function is parameterized by some decision variables and describes the mapping from sampling time to actual time. Then at each timestamp, VS-MPC solves an optimal control problem in which its decision variables consist of the control inputs in \eqref{oc} and the parameters of the time-warping function. With the situation changing at each timestamp, VS-MPC finds optimal control inputs and sampling settings jointly without manual tuning afterward.

\subsection{Time-warping Function}
\begin{figure}[h]
	\centering
	\includegraphics[width=0.30\textwidth]{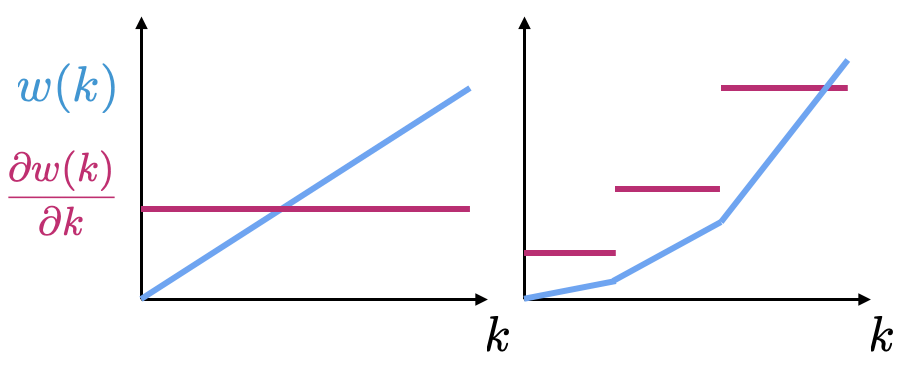}
	\caption{Two most common time warping functions for MPC}
	\label{fig:warping_fun_example}
\end{figure}

This paper proposes a differentiable time-warping function $w:\mathbb{R} \mapsto \mathbb{R}$ and denote $t = w(\tau)$, where $\tau \geq 0$ denotes the sampling time and $t \geq 0$ denotes the actual time. The states are sampled at $\tau = 0, 1, \cdots$ and the actual time associated with $\tau$ are $t = w(0), w(1), \cdots$. Some general constraints are considered on this time-warping function:
\begin{equation} \label{eq:time_warp_fun_require}
	w(0) = 0, \quad \frac{\partial w(\tau)}{\partial \tau}\Big|_{\tau=\hat{\tau}} > 0, \  \forall \hat{\tau} \geq 0
\end{equation}
Given the time-warping function $w(\cdot)$ and the formulation of \eqref{oc}, the time interval between two adjacent timestamp is
\begin{equation*}
	\Delta_j = w(j+1) - w(j).
\end{equation*}
Fig. \ref{fig:warping_fun_example} shows the two most common time warping functions for MPC. The left one indicates a uniform sampling when formulating an MPC problem, and the right one indicates a non-uniform sampling which partitions the entire horizon into multiple parts. Note that the right one is not differentiable at certain points.

To consider the system's future behavior when developing control, a single optimal control problem is often required to cover a total time $T$ given $N$ steps in the prediction horizon, i.e. $w(N) = T$. To make the prediction horizon adjustable, one can rewrite $w(N) = T$ as
\begin{equation} \label{eq:time_warp_fun_require:terminal}
	\underline{\alpha}T \leq w(N) \leq \overline{\alpha}T,
\end{equation}
where $\underline{\alpha}>0$ and $\overline{\alpha} > \underline{\alpha}$. Intuitively, one usually introduces small $\Delta_k$ in the near future and larger $\Delta_k$ in the distant future, which leads to one additional constraint:
\begin{equation} \label{eq:time_warp_fun_require:not_decreasing}
	\frac{\partial w(\tau)}{\partial \tau}\Big|_{\tau=\tau_2} - \frac{\partial w(\tau)}{\partial \tau}\Big|_{\tau=\tau_1} \geq 0, \ \forall \tau_2 \geq \tau_1 \geq 0.
\end{equation}

Without loss of generality, this paper chooses a polynomial with degree $s=2$ to represent a time-warping function:
\begin{equation*}
	t = \hat{w}(\tau, \boldsymbol{\beta}) = \beta_1 \tau + \beta_2 \tau^2,
\end{equation*}
where $\boldsymbol{\beta} = {\matt{\beta_1 & \beta_2}}^{\prime} \in \mathbb{R}^s$ is the coefficient vector that parameterizes the time warping function $\hat{w}$.
The constraints \eqref{eq:time_warp_fun_require} and \eqref{eq:time_warp_fun_require:not_decreasing} can be rewritten as
\begin{equation} \label{eq:time_warp_fun_require_poly}
	\begin{split}
		&\beta_1 + 2\beta_2 \tau > 0, \ \forall \tau \geq 0 \\
		&2\beta_2 \geq 0, \ \forall \tau \geq 0.
	\end{split}
\end{equation}

Thus the parameter $\boldsymbol{\beta}$ should satisfy the following conditions:
\begin{equation} \label{eq:time_warp_fun_require_poly:degree_3}
	\begin{split}
		&\underline{\alpha}T \leq \hat{w}(N, \boldsymbol{\beta}) \leq \overline{\alpha}T, \\
		&\beta_1 > 0, \ \beta_2 \geq 0.
	\end{split}
\end{equation}
Note that at time $t_k$, the current time warping function $\hat{w}_k(\tau, \boldsymbol{\beta}) = \hat{w}(\tau, \boldsymbol{\beta}) + t_k$ and $\frac{\partial \hat{w}_k(\tau, \boldsymbol{\beta})}{\partial \tau} = \frac{\partial \hat{w}(\tau, \boldsymbol{\beta})}{\partial \tau}$.

A parameterized time-warping function with a larger $s$ can represent a more complicated time mapping. However, it is impossible to construct constraints on $\boldsymbol{\beta}$ such that the conditions \eqref{eq:time_warp_fun_require} and \eqref{eq:time_warp_fun_require:not_decreasing} are satisfied when $s \geq 6$ because there is no algebraic solution to general polynomial equations of degree five or higher with arbitrary coefficients, as per Abel–Ruffini Theorem \cite{rosen1995niels}. Hence, one cannot obtain a constraint for $\boldsymbol{\beta} \in \mathbb{R}^s$ to satisfy \eqref{eq:time_warp_fun_require_poly} when $s \geq 6$.

Although a polynomial function is used to parameterize the
time-warping function, a time-warping function can be any function, as long as the constraints \eqref{eq:time_warp_fun_require}, \eqref{eq:time_warp_fun_require:terminal}, and \eqref{eq:time_warp_fun_require:not_decreasing} is satisfied. This paper uses polynomial parameterization due to its simplicity.



\subsection{Variable Sampling MPC}

If a non-uniform sampling is determined and well-tuned given the current situation, this sampling is not necessarily suitable for the next timestamp, which can be caused by external disturbance, etc.
Variable sampling MPC (VS-MPC) adapts how it samples along the horizon and determines optimal control accordingly at each timestamp $t_k$, without any offline manual tuning.

The VS-MPC strategy with variable sampling includes an optimal control problem at an arbitrary time $t_k$, which is formulated as follows:
\begin{mini}|s|
	{ \substack{ \boldsymbol{u}_{0:N-1|k} \\ \boldsymbol{\beta}_k \in \mathbb{R}^2 } }{ J(\boldsymbol{x}_{0:N|k},\boldsymbol{u}_{0:N-1|k}, \Delta_{0:N-1|k}) \label{prog:mpc_with_time_warping}}
	{}{}
	\addConstraint{ \boldsymbol{x}_{k+j+1|k} = \boldsymbol{x}_{k+j|k} + \Delta_{k+j|k} \boldsymbol{f}_c(\boldsymbol{x}_{k+j|k}, \boldsymbol{u}_{k+j|k})}
	\addConstraint{\forall j=0,\cdots,N-1 \  \text{with given } \boldsymbol{x}_{k}}
	\addConstraint{\boldsymbol{g}(\boldsymbol{x}_{0:N|k}, \boldsymbol{u}_{0:N-1|k}, \Delta_{0:N-1|k}) \leq \boldsymbol{0}}
	\addConstraint{\boldsymbol{h}(\boldsymbol{x}_{0:N|k}, \boldsymbol{u}_{0:N-1|k}, \Delta_{0:N-1|k}) = \boldsymbol{0}}
	\addConstraint{\Delta_{k+j|k} = \hat{w}(j+1, \boldsymbol{\beta}_k) - \hat{w}(j, \boldsymbol{\beta}_k)}
	\addConstraint{\forall j=0, \cdots, N-1}
	\addConstraint{\underline{\alpha}T \leq \hat{w}(N,\boldsymbol{\beta}_k) \leq \overline{\alpha}T}
	\addConstraint{-\beta_1 <0, \  -\beta_2 \leq 0.}
\end{mini}
At each timestamp $t_k$, VS-MPC determines the optimal control and how it samples given \eqref{prog:mpc_with_time_warping}. Then it performs the optimal control and repeats this process in the next timestamp.
The detailed explanation of the algorithm for VS-MPC is shown below, where $\frac{1}{dt_{mpc}}$ indicates the MPC frequency. The MPC frequency is typically determined by the specifications of the actual controller.
\begin{algorithm}
	\caption{Variable Sampling MPC}\label{alg:vs_mpc}
	\DontPrintSemicolon
	\KwIn{$k = 0$, $t_0$, $x(t_0)$, $N$, $dt_{mpc}>0$}
	\While {true} {
		$\boldsymbol{u}^*_{0:N-1|k}, \ \boldsymbol{\beta}^*_k \gets$  solve $\eqref{prog:mpc_with_time_warping}$\;
		
		$\hat{w}_k(\cdot, \boldsymbol{\beta}) = \hat{w}(\cdot, \boldsymbol{\beta}) + t_k$\;
		
		$\boldsymbol{u}_k \gets$ interpolate $\boldsymbol{u}^*_{0:N-1|k}$ for time $[t_k,t_k + dt_{mpc}]$ given $\hat{w}_k(\cdot, \boldsymbol{\beta})$ and zero-order hold\;
		perform $\boldsymbol{u}_k$ until $t_k + dt_{mpc}$\;
		$t_{k} \gets t_k + dt_{mpc}$\;
		$k \gets k+1$\;
	}
\end{algorithm}

\section{Simulations}\label{sec:simulation}

This section shows how the VS-MPC strategy is applied to a battery energy storage system (BESS) for a wind farm and compares the revenue regarding two MPC strategies with uniform sampling and variable sampling.

\subsection{Battery Energy Storage System for Wind Farm}

\begin{figure}[h]
	\centering
	\includegraphics[width=0.30\textwidth]{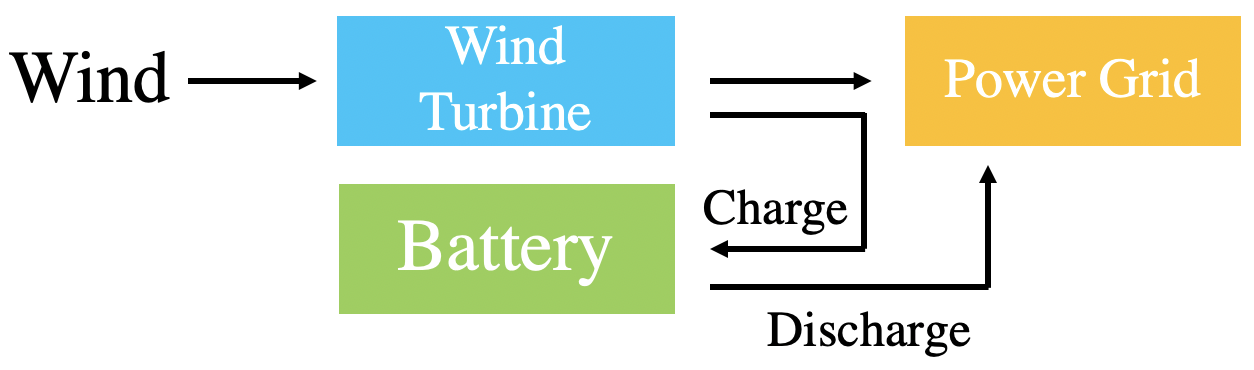}
	\caption{A battery energy storage system for a wind farm}
	\label{fig:bess_example}
\end{figure}

This subsection discusses how to design controls of a BESS by MPC to provide reserves to mitigate wind power intermittency. The following problem formulation originates from Ref. \cite{li2013mpc}.
In particular, an MPC strategy will be used to control the charge and discharge of the battery in BESS to reduce the negative impact caused by wind intermittency. As shown in Fig. \ref{fig:bess_example}, the MPC strategy is required to decide, at each timestamp, how much wind power goes to the power grid and how much goes to the BESS or how much power the BESS discharge and then goes to the grid.

Assume that the nameplate capacity [MWh] of the wind farm is denoted by $Q_n > 0$. Assuming that the efficiencies of both charge and discharge are perfect, and the responses of both are instantaneously fast, the dynamics of the battery state of charge (SOC) are governed as follows:
\begin{equation} \label{example:battery_dynamics}
	x(t) = -\frac{P_{batt}(t)}{Q_c},
\end{equation}
where $x(t) \in [0,1]$ is the battery SOC at time $t$; $P_{batt}(t) \in \mathbb{R}$ is the battery discharge power [MW] and $P_{batt}(t) < 0$ indicates the battery is charging at time $t$; $Q_c > 0$ is the battery capacity [MWh].

Suppose that the control input is the scheduling wind power $u(t) \in \mathbb{R}_{\geq 0}$ at time $t$ that goes to the grid and denotes the actual wind power at time $t$ as $w_a(t) \in \mathbb{R}_{\geq 0}$, then $P_{batt}(t) = u(t) - w_a(t)$. The system dynamics of the BESS are written as follows:
\begin{equation} \label{example:bess_dynamics}
	\dot{x}(t) = f_c(x(t), u(t)) = \frac{w_a(t)-u(t)}{Q_c}.
\end{equation}
Assume that $w_a(t)$ is unknown before determining the power scheduling $u(t)$ at time $t$ but there is a wind power forecasting $w_f(t) \in \mathbb{R}_{\geq 0}$ available at time $t$. Then, one needs to determine the power scheduling based on wind power forecasting.


The cost function includes the revenue of selling wind power to the grid, the expense of scheduling conventional reserves based on wind forecasting, the expense of dispatching conventional reserves due to the mismatch between actual and forecasted wind power, and the expense of ramping services\cite{li2013mpc}.
Denote the power reserve requirement at time $t_k$ as $r(u,w_f) = [u(t_k)-w_f(t_k)]^+$ and the wind power shortage due to the imperfect forecasting at time $t_k$ as $d(u,w_a)=[u(t_k)-w_a(t_k)]^+$.
Denote $\overline{P}(x_k) \in \mathbb{R}_{\geq 0}$ as the battery discharging power limit given the SOC $x_k$ at time $t_k$. Similarly, $\underline{P}(x_k) \in \mathbb{R}_{\leq 0}$ denotes the battery charging power limit.
Then the cost $c_k$ at time $t_k$ is defined by
\begin{equation}
	\begin{split}
		c_k = &-\alpha_1 u_{k} + \alpha_2\Big[r\big(u_{k},w_f(t_k)\big)-\overline{P}(x_k)\Big]^+ \\
		&+\alpha_3\Big[d\big(u_{k},w_a(t_k)\big)-\overline{P}(x_k)\Big]^+\\
		&+\alpha_4|u_k-u_{k-1}|.
	\end{split}
\end{equation}
Note that $c_k$ cannot be evaluated at current time $t_k$ because $w_a(t_k)$ is unknown. Instead, the estimated cost $\hat{c}_{k+j|k}$ of time $t_{k+j}$ that is predicted at time $t_k$ is defined as
\begin{equation}
	\begin{split}
		\hat{c}_{k+j|k} = &-\alpha_1 u_{k+j|k} + (\alpha_2+\alpha_3)\Big[r\big(u_{k+j|k},w_f(t_{j+k})\big)\\
		&-\overline{P}(x_{k+j|k})\Big]^+ +\alpha_4\ell(u_{k+j|k}-u_{k+j-1|k}),
	\end{split}
\end{equation}
where $\ell:\mathbb{R} \mapsto \mathbb{R}_{> 0}$ denotes a smooth approximation of absolute value function $|\cdot|$ and $\ell(x)=\sqrt{x^2+0.01}$. The coefficients $\alpha_1, \alpha_2, \alpha_3$, and $\alpha_4$ are the unit price of electricity generation, reserve scheduling, reserve dispatch, and ramping services in the wholesale market, respectively. And these coefficients are determined based on statistics in \cite{monitor20112010}, i.e.
\begin{equation}
	\alpha_1=1, \alpha_2=1.03, \alpha_3=1, \alpha_4=0.5455.
\end{equation}
Thus, the total estimated cost over time horizon $[t_k, t_{k+N}]$ is defined by
\begin{equation} \label{example:objective_func}
	\begin{split}
		\hat{J}_k= \textstyle\sum_{j=0}^{N-1} \hat{c}_{k+j|k}\Delta_{k+j|k},
	\end{split}
\end{equation}
where $u_{k-1|k}$ indicates the previous control input at time $t_{k-1}$.
Hence, the optimal control problem with variable sampling \eqref{prog:mpc_with_time_warping} at each time $t_k$ can be rewritten as follows:
\begin{mini}|s|
	{ \substack{ u_{0:N-1|k} \\ \boldsymbol{\beta}_k \in \mathbb{R}^2 } }{ \frac{\hat{J}_k}{w(N, \boldsymbol{\beta})} \label{prog:batt:mpc_with_time_warping_final}}
	{}{}
	\addConstraint{ x_{k+j+1|k} = x_{k+j|k} + \Delta_{k+j|k} f_c(x_{k+j|k}, u_{k+j|k})}
	\addConstraint{\Delta_{k+j|k} = w(j+1, \boldsymbol{\beta}_k) - w(j, \boldsymbol{\beta}_k)}
	\addConstraint{\underline{\alpha}T \leq w(N,\boldsymbol{\beta}_k) \leq \overline{\alpha}T}
	\addConstraint{-\beta_1 <0, \  -\beta_2 \leq 0.}
	\addConstraint{SOC_{min} \leq x_{k+j|k} \leq SOC_{max}}
	\addConstraint{0 \leq u_{k+j|k} \leq Q_n}
	\addConstraint{\underline{P}(x_{k+j|k}) \leq u_{k+j|k} - w_f(t_{k+j}) \leq \overline{P}(x_{k+j|k})}
	\addConstraint{\forall j=0,\cdots,N-1 \  \text{with given } x_{k},}
\end{mini}
where the objective $\frac{\hat{J}_k}{w(N, \boldsymbol{\beta})}$ indicates the average cost over the entire prediction horizon $[0, \  w(N,\boldsymbol{\beta})]$.

\subsection{Result}
The parameters are: $N=10$, $T=1$ hour, $Q_n=400$ MWh, $\underline{\alpha}=1$, $\overline{\alpha}=4$, $SOC_{min}=0.3$, $SOC_{max}=0.9$, $x(0)=0.4$. When $Q_c \leq Q_n$, the battery charge and discharge limits are defined as follows \cite{li2013mpc}:
\begin{equation}
	\overline{P}(x) = Q_c x, \ \underline{P}(x) = Q_c(x-1), \ x \in [0,1].
\end{equation}
When $Q_c > Q_n$, the limits are defined by
\begin{equation}
	\overline{P}(x)=\begin{cases}
		Q_c x, & \text{$x \in [0, \ \frac{Q_n}{Q_c}]$}\\
		Q_n, & \text{$x \in [\frac{Q_n}{Q_c}, \ 1]$}
	\end{cases},
\end{equation}
and
\begin{equation}
	\underline{P}(x)=\begin{cases}
		-Q_n, & \text{$x \in [0, \ 1-\frac{Q_n}{Q_c}]$}\\
		Q_c(x-1), & \text{$x \in [1-\frac{Q_n}{Q_c}, \ 1]$}
	\end{cases}.
\end{equation}
\begin{figure}[h]
	\centering
	\includegraphics[width=0.40\textwidth]{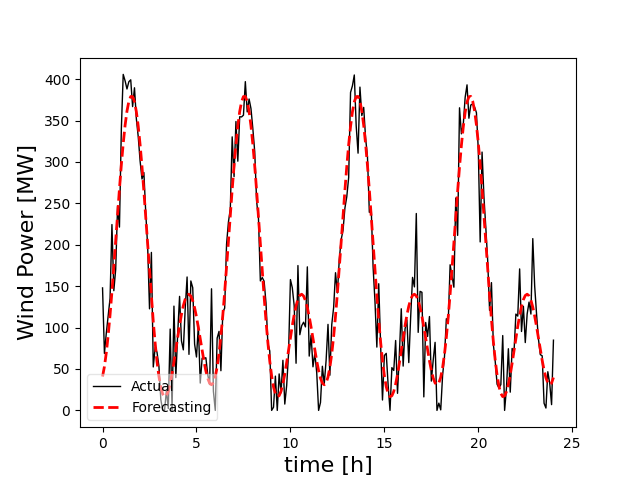}
	\caption{The actual and forecasting wind power trajectories. $w_f(t) = 120\text{sin}(\frac{\pi t}{3})+100\text{sin}(2\pi \frac{t+2}{3} + 0.4)+150$. The actual wind $w_a(t)$ equals to the forecasting $w_t(t)$ adding a random noise whose distribution is $\mathcal{N}(0, 40^2)$. The actual wind is clipped to zero when negative.}
	\label{fig:wind_traj}
\end{figure}
The simulation's time step is 0.1 hour and the entire simulation horizon is 24 hours. The wind trajectories are shown in Fig. \ref{fig:wind_traj}.
Ref. \cite{li2013mpc} also proposes a heuristic control algorithm, which is written as follows:
\begin{equation}
	u(t) = w_f(t)\cdot 2x(t).
\end{equation}
This section adopts both the heuristic control algorithm and an MPC strategy with uniform sampling for revenue comparisons. The prediction horizon for the MPC with uniform sampling is 1 hour and includes 10 steps.

\begin{figure}[h]
	\centering
	\includegraphics[width=0.40\textwidth]{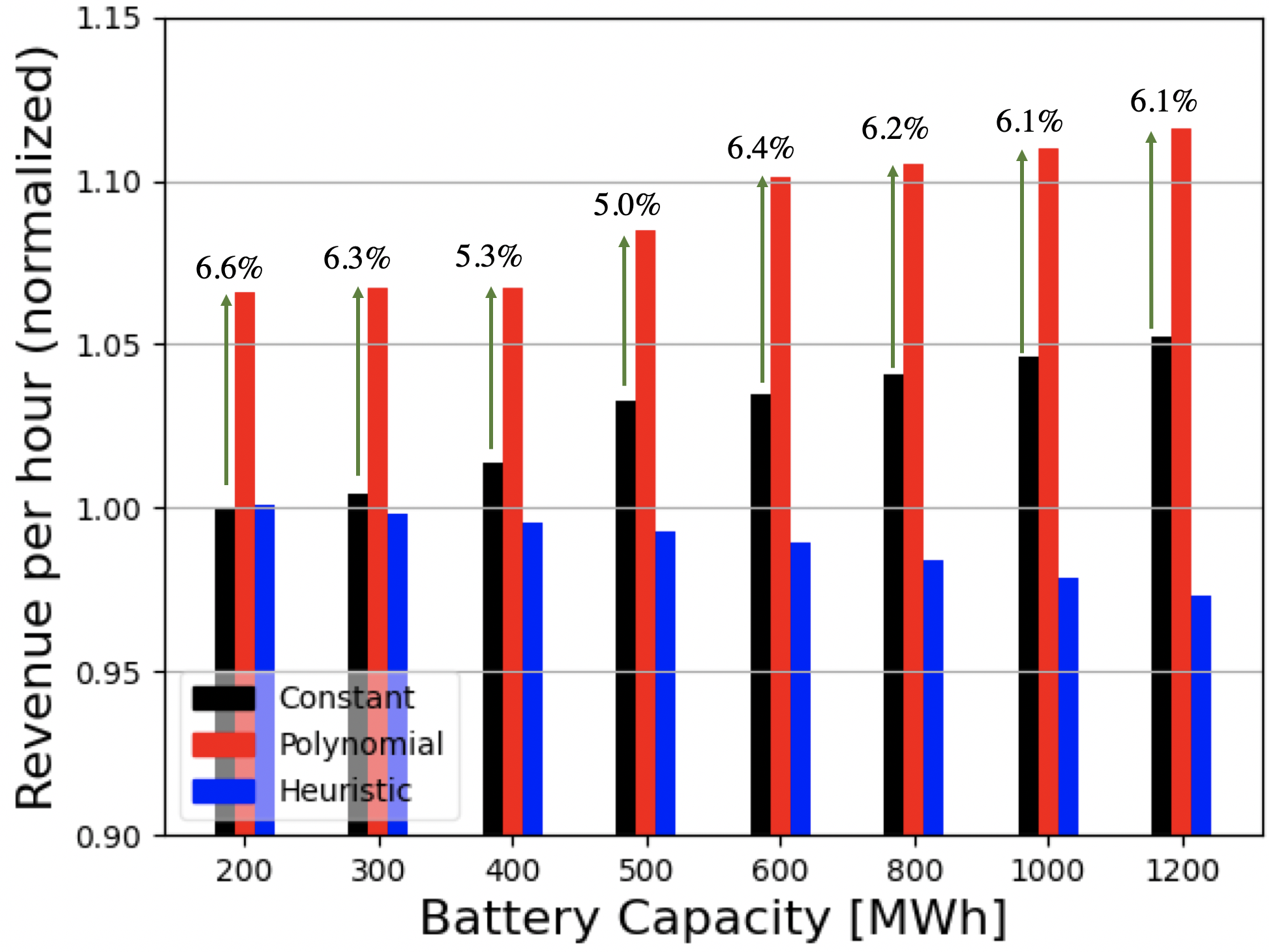}
	\caption{Average revenue (normalized) when wind forecasting is perfect. The number over the red bar is the relative difference in revenue between the two MPC strategies.
	}
	\label{fig:comp:wind_perfect}
\end{figure}

Fig. \ref{fig:comp:wind_perfect} compares the average revenue (the converse of the total cost) given different battery capacities and control strategies when wind forecasting is perfect. The battery capacity varies from 200 MWh (50\% of the nameplate) to 1200 MWh (300\%). And all the average revenues are normalized as the revenue of 200 MWh given uniform sampling MPC is set to 1. Fig. \ref{fig:comp:wind_perfect} shows that the proposed VS-MPC strategy outperforms the other two methods for all battery capacities listed in the figure. As the battery capacity increases, the revenue for two MPC strategies increases because a battery with a larger capacity is more capable of compensating for wind intermittency.
The revenue difference between the two MPC strategies is roughly constant as the battery capacity grows. Since the maximum length of the prediction horizon for VS-MPC is fixed over these cases, the exclusive look-ahead information that VS-MPC obtains is nearly the same. 
And the revenue difference between the heuristic control algorithm and MPC strategies grows when the battery capacity rises because the heuristic algorithm does not exploit any look-ahead information.

\begin{figure}[h]
	\centering
	\includegraphics[width=0.40\textwidth]{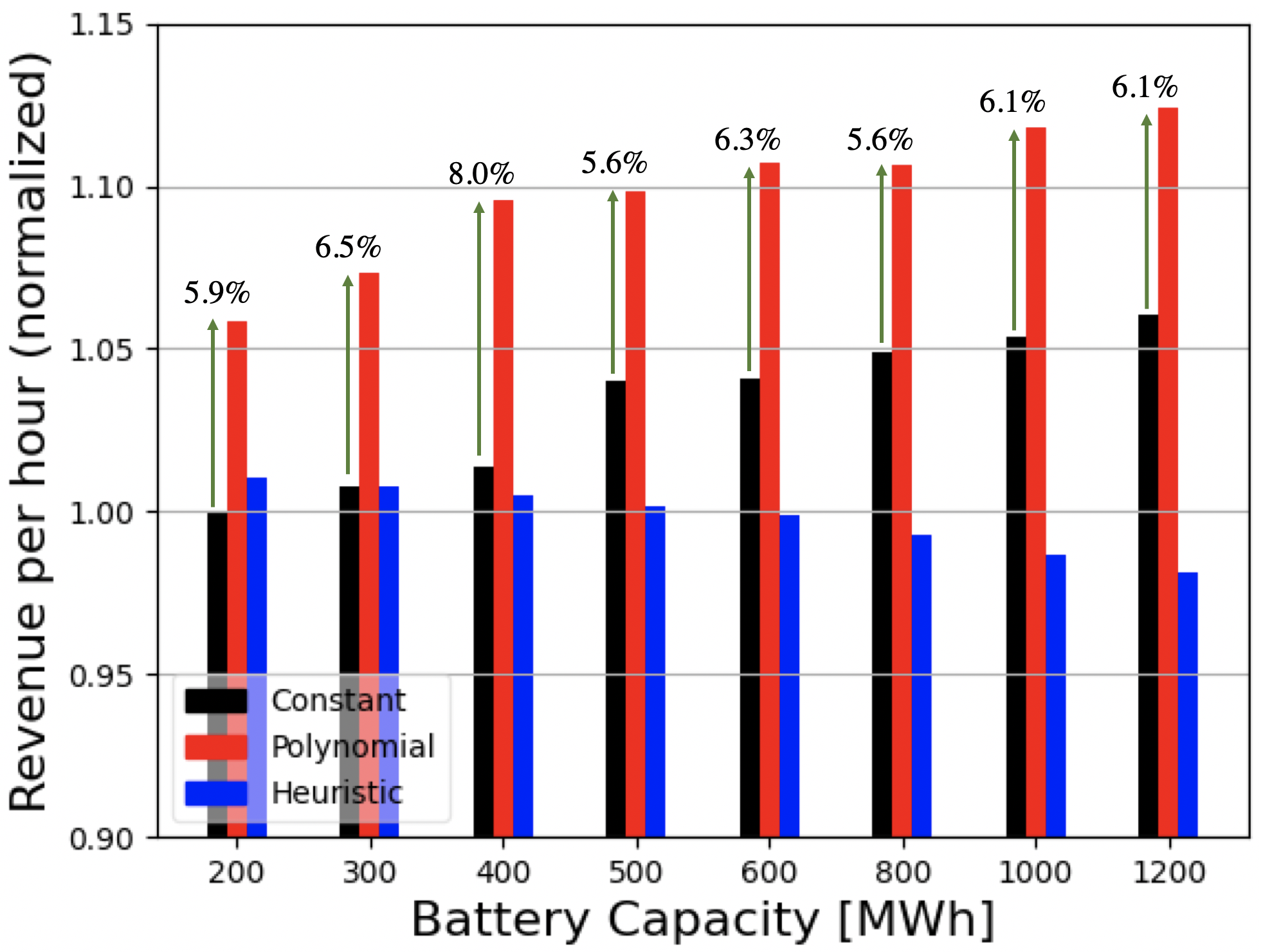}
	\caption{Average revenue (normalized) when wind forecasting is imperfect}
	\label{fig:comp:wind_guassian}
\end{figure}

Fig. \ref{fig:comp:wind_guassian} compares the average revenue given different battery sizes and control strategies when wind forecasting is imperfect (wind trajectories shown in Fig. \ref{fig:wind_traj}). Fig. \ref{fig:comp:wind_guassian} also reveals that the VS-MPC strategy outperforms the other two methods for all battery capacities. And all the other observations are consistent with those mentioned in Fig. \ref{fig:comp:wind_perfect}.

\begin{figure}[h]
	\centering
	\begin{subfigure}[b]{0.35\textwidth}
		\includegraphics[width=\textwidth]{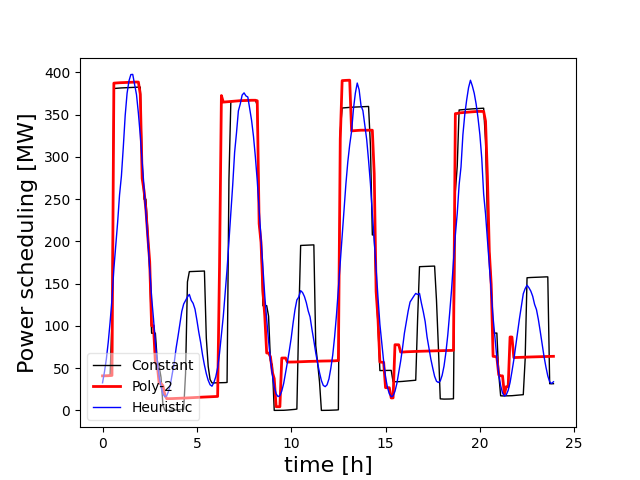}
		\caption{The trajectories of input for 3 methods.}
		\label{fig:example_f2_400:input}
	\end{subfigure}
	\hfill
	\centering
	\begin{subfigure}[b]{0.35\textwidth}
		\includegraphics[width=\textwidth]{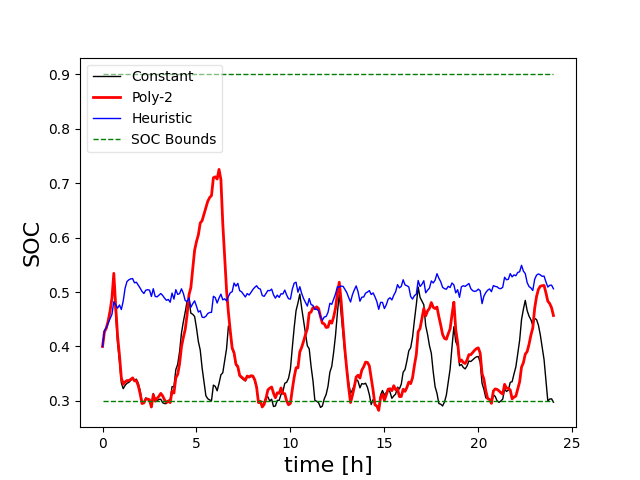}
		\caption{The trajectories of SOC for 3 methods.}
		\label{fig:example_f2_400:soc}
	\end{subfigure}
	\caption{Details about the BESS system when the wind forecasting is imperfect. $Q_c$ = 400 MWh.}
	\label{fig:example_f2_400}
\end{figure}

Fig. \ref{fig:example_f2_400} shows the trajectories of the control input (power scheduling) and the state (SOC) for 3 methods when $Q_c$ = 400 MWh. Since the heuristic algorithm does not use any prediction, its SOC oscillates around 0.5 due to the forecasting error. Since it does not use much battery capacity for compensating wind intermittency, its average revenue underperforms the MPC strategies. As for the VS-MPC, since it is possible to exploit much more look-ahead information by a longer horizon, its behavior around 5, 10, 17, and 22 hours is more smooth than the uniform sampling MPC, which reduces the ramping cost.

\section{Conclusion}\label{sec:conclusion}
This paper proposes a variable sampling model predictive control (VS-MPC) strategy, which can deal with multi-timescale systems with only one controller. Unlike the existing non-uniform sampling MPC (NS-MPC) or multi-horizon MPC (MH-MPC) strategies, VS-MPC does not require offline and manual tuning on some parameters for the prediction horizon. Instead, VS-MPC constructs a differentiable and parameterized time-warping function to describe the sampling nature of a non-uniform horizon. Then an optimization program jointly determines the optimal control inputs and the parameters for the time-warping function at each timestamp. Lastly, this paper uses an example of BESS for a wind farm to demonstrate the performance of VS-MPC. Some revenue comparisons for several methods have been provided to show the advantages of the proposed VS-MPC. Future work includes an extension of the proposed VS-MPC to be tunable concerning additional loss or constraints \cite{Mou2020NIPS, Mou2021NIPS}, cooperative tuning of VS-MPC for multi-agent systems \cite{Zehui22CDC}, and application of VS-MPC into other practical systems such as battery management systems \cite{Huazhen21Battery}.

\bibliographystyle{ieeetr}
\bibliography{Reference}

\end{document}